\documentclass[letterpaper, 10 pt, conference]{ieeeconf} 
\IEEEoverridecommandlockouts   

\usepackage{tikz}
\usetikzlibrary{arrows.meta,positioning,shapes.geometric}
\usepackage{amsmath,amssymb,amsfonts}
\usepackage{subcaption}
\usepackage{float}

\usepackage{comment}
\usepackage{algorithm}
\usepackage{algpseudocode}
\usepackage{graphicx}
\usepackage{textcomp}
\usepackage{xcolor}
\usepackage{booktabs}
\usepackage{multirow}
\usepackage{optidef}
\usepackage{tikz}
\usepackage{enumerate}
\usetikzlibrary{arrows.meta,positioning,calc,backgrounds}

\tikzset{
  mbox/.style={draw=black!70,fill=white,rounded corners=3pt,
    line width=0.9pt,inner sep=0pt},
  sbox/.style={draw=black!40,fill=black!4,rounded corners=2pt,
    line width=0.6pt,inner sep=5pt,align=center,
    font=\scriptsize\sffamily},
  ltitle/.style={font=\small\sffamily\bfseries,text=black},
  slabel/.style={font=\scriptsize\sffamily,text=black!60},
  tlabel/.style={font=\fontsize{6.5}{8}\selectfont\sffamily,text=black!50},
  sig/.style={-{Stealth[length=4.5pt,width=3.5pt]},
    line width=0.9pt,color=black},
  sigdash/.style={sig=black,dashed,dash pattern=on 3pt off 2pt},
}

\newtheorem{remark}{Remark}

\newcommand{\ying}[1]{\textcolor{black}{#1}}

\begin{document}

\title{Decentralized Opinion-Integrated Decision making 
at Unsignalized Intersections via Signed Networks }

\author{Bhaskar Varma$^{*}$, Ying Shuai Quan$^{**}$, Karl D. von Ellenrieder$^{*}$ and Paolo Falcone$^{**}$%
\thanks{$^{*}$Bhaskar Varma and Karl D. von Ellenrieder are with the Faculty of Engineering, Free University of Bozen-Bolzano, Italy
{\tt\footnotesize \{sbalagopala, karl.vonellenrieder\}@unibz.it}.
$^{**}$Ying Shuai Quan and Paolo Falcone are with the Mechatronics Group, Department of Electrical Engineering, Chalmers University of Technology, Gothenburg, Sweden
{\tt\footnotesize \{quany,paolo.falcone\}@chalmers.se}.}
}

\maketitle

\begin{abstract}
\begin{abstract}
In this letter, we consider the problem of decentralized 
decision making among connected autonomous vehicles at 
unsignalized intersections, where existing centralized 
approaches do not scale gracefully under mixed maneuver 
intentions and coordinator failure. We propose a closed-loop 
opinion-dynamic decision model for intersection coordination, 
where vehicles exchange intent through dual signed networks: 
a conflict topology based communication network and a 
commitment-driven belief network, enable cooperation 
without a centralized coordinator. Continuous opinion states 
modulate velocity optimizer weights prior to commitment; a 
closed-form predictive feasibility gate then freezes each 
vehicle's decision into a \textsc{go} or \textsc{yield} 
commitment, which propagates back through the belief network 
to pre-condition neighbor behavior ahead of physical 
conflicts. Crossing order emerges from geometric feasibility 
and arrival priority without the use of joint optimization or a solver. 
The approach is validated across \ying{three} scenarios spanning 
\ying{fully competitive, merge, and mixed} conflict 
topologies\ying{. The results demonstrate collision-free coordination 
and lower last-vehicle exit times compared to first come first served (FCFS) in all 
conflict-non-trivial configurations}.
\end{abstract}

\end{abstract}
\section{Introduction and Related Work}
\label{sec:intro}

Urban unsignalized intersections represent one of the most demanding 
coordination environments for connected autonomous vehicles.
While the problem is not new and looks solved using a centralized 
model predictive control (MPC) like~\cite{cmpc}, the main issue is 
that it relies on a centralized coordinator and considers only 
longitudinal control along straight paths, a limitation that persists 
even in decentralized MPC extensions~\cite{9745747}. \ying{Reservation-based 
methods~\cite{10452809} treat} the intersection as a shared resource 
and assign time slots through a centralized intersection manager. 
\ying{Some} intersection managers propose~\cite{8603190} a First Come 
First Served (FCFS) policy instead of an optimal passing order; 
\ying{this works adequately} if you only consider straight\ying{-path} 
\ying{maneuvers}, but with turn maneuvers the overall throughput will 
be compromised. Recent approaches \ying{take} the lane geometry and 
turns into consideration~\cite{lin2025safety}, but again reintroduce a 
centralized controller, which \ying{increases} the computational 
complexity and dependence when upscaled. Also, learning-based 
optimization algorithms \ying{have been proposed}~\cite{gong2024collision} 
yet remain dependent on offline training and generalize poorly across 
unseen configurations. These approaches either require a persistent 
central authority that fails to scale gracefully under coordinator 
failure or intermittent communication, or rely on optimization-based 
joint and distributed MPC formulations that achieve strong performance 
at the cost of computational complexity.

On the other hand, opinion dynamics~\cite{9736598} \ying{have recently 
attracted attention} especially for decentralized\ying{,} 
\ying{interactive} decision making\ying{:} \ying{including} robots 
interacting with human movers in~\cite{10341745} \ying{and for} task 
allocation in uncrewed surface vessels~\cite{paine2024model}, and 
\ying{have} also been used in unsignalized roundabouts~\cite{liu2026cooperative}. 
Opinion dynamics on signed social networks~\cite{zaslavsky2013matrices}
offer a different paradigm: distributed agents form opinions through
local interactions, and bifurcation phenomena naturally produce
differentiated equilibria without explicit negotiation. 
\ying{The natural bifurcation of signed-network equilibria into two 
opposing committed states maps directly onto the binary GO/YIELD 
decision required at intersections.} 
The multitopic extension in~\cite{10874167} establishes conditions 
under which structurally balanced signed networks produce stable 
committed equilibria, providing theoretical grounding for 
opinion-based coordination.

\ying{The primary focus of this work is to study interactive decision 
making at intersections, not throughput maximization as in high-density 
traffic management.} While extensive literature addresses the latter, 
the core problem of individual vehicle decision-making is often 
bypassed through centralized sequencing, rather than vehicles resolving 
it distributively. We address this gap with a decentralized architecture 
integrating opinion dynamics and decomposed predictive control\ying{,
where a closed-form feasibility gate freezes each vehicle's commitment.
Opinion evolution over dual signed networks facilitates continuous
coordination without a central authority.}
This letter makes the following contributions:

\begin{enumerate}

\item \textbf{Dual signed graph architecture.}
A conflict based communication graph and commitment driven 
belief network together define structured Suppression, Permission, 
and Coordination influence, routing committed state through opinion
dynamics without a weighted coupling matrix.

\item \textbf{Decomposed predictive coordination.}
MPC is decomposed into a closed-form predictive
feasibility check, which drives a decentralized commitment
gate and a single-step multi-objective velocity
optimizer, while preserving predictive safety.

\item \textbf{Opinion-integrated control.}
The continuous opinion modulates optimizer weights before commitment; 
each commitment event updates the belief network, \ying{triggering 
opinion spikes that propagate intent through the network ahead of 
physical conflict resolution.}

\end{enumerate}

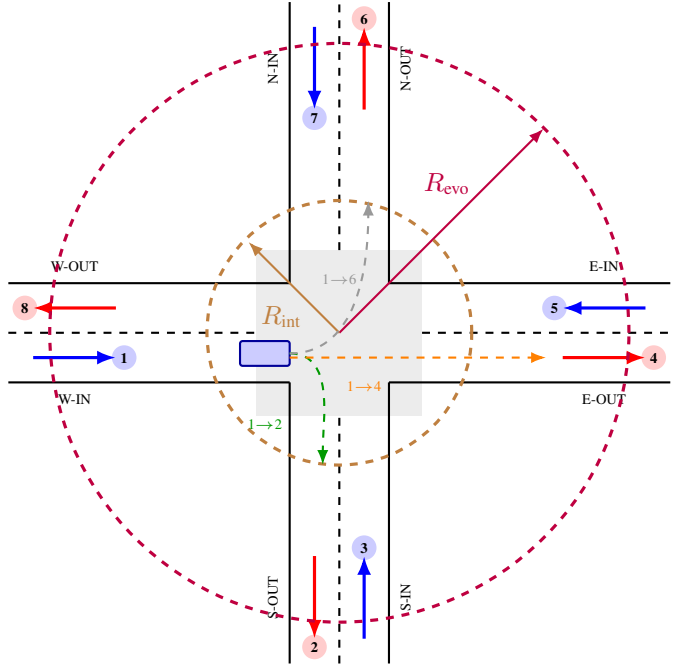
\begin{figure}[!b]
\centering
\begin{tikzpicture}[scale=1.1, every node/.style={scale=1.1}]
\fill[gray!15] (-1,-1) rectangle (1,1);
\fill[white] (-4,-0.6) rectangle (-2,0.6);
\draw[thick] (-4,0.6) -- (-0.6,0.6);
\draw[thick] (-4,-0.6) -- (-0.6,-0.6);
\draw[dashed,black,thick] (-4,0) -- (-1,0);
\fill[white] (2,-0.6) rectangle (4,0.6);
\draw[thick] (0.6,0.6) -- (4,0.6);
\draw[thick] (0.6,-0.6) -- (4,-0.6);
\draw[dashed,black,thick] (1,0) -- (4,0);
\fill[white] (-0.6,-4) rectangle (0.6,-2);
\draw[thick] (-0.6,-4) -- (-0.6,-0.6);
\draw[thick] (0.6,-4) -- (0.6,-0.6);
\draw[dashed,black,thick] (0,-4) -- (0,-1);
\fill[white] (-0.6,2) rectangle (0.6,4);
\draw[thick] (-0.6,0.6) -- (-0.6,4);
\draw[thick] (0.6,0.6) -- (0.6,4);
\draw[dashed,black,thick] (0,1) -- (0,4);
\node[fill=blue!20,circle,inner sep=1.5pt] at (-2.6,-0.3) {\tiny\textbf{1}};
\draw[-{Latex[length=2.5mm]},blue,very thick] (-3.7,-0.3) -- (-2.7,-0.3);
\node[font=\tiny] at (-3.2,-0.8) {W-IN};
\node[fill=red!20,circle,inner sep=1.5pt] at (-3.8,0.3) {\tiny\textbf{8}};
\draw[-{Latex[length=2.5mm]},red,very thick] (-2.7,0.3) -- (-3.7,0.3);
\node[font=\tiny] at (-3.2,0.8) {W-OUT};
\node[fill=blue!20,circle,inner sep=1.5pt] at (2.6,0.3) {\tiny\textbf{5}};
\draw[-{Latex[length=2.5mm]},blue,very thick] (3.7,0.3) -- (2.7,0.3);
\node[font=\tiny] at (3.2,0.8) {E-IN};
\node[fill=red!20,circle,inner sep=1.5pt] at (3.8,-0.3) {\tiny\textbf{4}};
\draw[-{Latex[length=2.5mm]},red,very thick] (2.7,-0.3) -- (3.7,-0.3);
\node[font=\tiny] at (3.2,-0.8) {E-OUT};
\node[fill=blue!20,circle,inner sep=1.5pt] at (0.3,-2.6) {\tiny\textbf{3}};
\draw[-{Latex[length=2.5mm]},blue,very thick] (0.3,-3.7) -- (0.3,-2.7);
\node[font=\tiny,rotate=90] at (0.8,-3.2) {S-IN};
\node[fill=red!20,circle,inner sep=1.5pt] at (-0.3,-3.8) {\tiny\textbf{2}};
\draw[-{Latex[length=2.5mm]},red,very thick] (-0.3,-2.7) -- (-0.3,-3.7);
\node[font=\tiny,rotate=90] at (-0.8,-3.2) {S-OUT};
\node[fill=blue!20,circle,inner sep=1.5pt] at (-0.3,2.6) {\tiny\textbf{7}};
\draw[-{Latex[length=2.5mm]},blue,very thick] (-0.3,3.7) -- (-0.3,2.7);
\node[font=\tiny,rotate=90] at (-0.8,3.2) {N-IN};
\node[fill=red!20,circle,inner sep=1.5pt] at (0.3,3.8) {\tiny\textbf{6}};
\draw[-{Latex[length=2.5mm]},red,very thick] (0.3,2.7) -- (0.3,3.7);
\node[font=\tiny,rotate=90] at (0.8,3.2) {N-OUT};
\draw[dashed,purple,very thick] (0,0) circle (3.5);
\draw[-{Latex[length=2mm]},purple,thick] (0,0) -- (2.47,2.47);
\node[purple,font=\small] at (1.3,1.8) {$R_{\text{evo}}$};
\draw[dashed,brown,very thick] (0,0) circle (1.6);
\draw[-{Latex[length=2mm]},brown,thick] (0,0) -- (-1.1,1.1);
\node[brown,font=\small] at (-0.7,0.2) {$R_{\text{int}}$};
\draw[fill=blue!20, draw=blue!60!black, thick, rounded corners=1pt]
    (-1.2,-0.4) rectangle (-0.6,-0.1);
\draw[-{Latex[length=2mm]},green!60!black,thick,dashed] (-0.6,-0.25) to[out=0,in=90] (-0.2,-1.6);
\node[green!60!black,font=\tiny] at (-0.9,-1.1) {1$\to$2};
\draw[-{Latex[length=2mm]},orange,thick,dashed] (-0.6,-0.3) -- (2.5,-0.3);
\node[orange,font=\tiny] at (0.3,-0.65) {1$\to$4};
\draw[-{Latex[length=2mm]},white!60!black,thick,dashed] (-0.6,-0.25) to[out=0,in=270] (0.35,1.6);
\node[white!60!black,font=\tiny] at (0,0.65) {1$\to$6};
\end{tikzpicture}
\caption{Four-way unsignalized intersection. In-lanes (blue, labeled 1,3,5,7)
and out-lanes (red, labeled 2,4,6,8). Maneuvers from lane~1:
right turn 1$\to$2 (green), straight 1$\to$4 (orange), left turn 1$\to$6
(gray). Red circle:
evolution zone $R_{\text{evo}}=15$\,m }
\label{fig:intersection}
\vspace{-0.3cm}
\end{figure}
\section{Problem Statement}
\label{sec:problem}

\subsection{Intersection Geometry}

We consider a four-way unsignalized intersection with right-hand traffic
rules as shown in Fig.~\ref{fig:intersection}. Each approach (West, South,
East, North) has two lanes: an in-lane for approaching vehicles and
an out-lane for departing vehicles. In-lanes are labeled $1, 3, 5, 7$
(West, South, East, North) and out-lanes $2, 4, 6, 8$ respectively.
\subsection{Problem Statement}

Given a set of CAVs approaching a four-way unsignalized intersection
with mixed maneuver intentions, design a distributed coordination
system such that:
\begin{enumerate}[(1)]
    \item \textbf{Safety:} all vehicles complete their intended
          maneuvers without collision;
          \item\textbf{Decentralization:} without central coordinator or
          pre-assigned crossing order.
\end{enumerate}

\subsection{Vehicle Model and Path Execution}

 Consider $N_a$ CAVs with state $\mathbf{x}_i = [x_i, y_i,
\theta_i, v_i]^\top$. Each vehicle follows a pre-computed arc-length
parameterized path determined by its declared (in-lane, out-lane) intention.
Path tracking is assumed ideal: lateral position is read directly from the
path. 
The coordination layer determines
$\dot{v}_i = a_i(t)$; path geometry serves as a  fixed spatial reference along which each vehicle  progresses.The length of each vehicle $L_v$ is  4.5m and the width is 1.8m for the simulations.


\section{Method}
\label{sec:method}

\subsection{Signed Opinion Dynamics Model}

We build on the nonlinear opinion dynamics framework of
\cite{10874167}. The $N_a$ agents are on a signed network
with adjacency matrix $A \in \mathbb{R}^{N_a \times N_a}$,
where $a_{ik} > 0$ denotes cooperative and $a_{ik} < 0$
antagonistic interactions between \ying{agents $i$ and $k$}. Let $z_{ij} \in \mathbb{R}$ be
the opinion of agent $i$ about option $j$. The
continuous-time dynamics are
\begin{equation}
\tau\,\dot{z}_{ij} = -d\,z_{ij} +
\tanh\!\left(u\,I_{ij} + b_{ij}\right)
\label{eq:base_nod},
\end{equation}
where the internal state is
\begin{align}
I_{ij} = \;&\alpha\, z_{ij}
+ \gamma \sum_{k \neq i} a_{ik}\, z_{kj} \nonumber\\
&+ \beta \sum_{l \neq j} g_{jl}\, z_{il}
+ \delta \sum_{k \neq i}\sum_{l \neq j}
  a_{ik}\, g_{jl}\, z_{kl}.
\label{eq:internal}
\end{align}
Here $d{>}0$ is resistance, $u{\geq}0$ is attention,
$b_{ij}$ is external bias, and $g_{jl}$ are entries of the
belief-system graph $G$. The four governing parameters are:
$\alpha$ (self-reinforcement), $\gamma$ (inter-agent
coupling via signed $A$), $\beta$ (cross-option coupling
via $G$), and $\delta$ (combined inter-agent and
cross-option). The neutral state $Z{=}0$ undergoes a
pitchfork bifurcation at a critical attention $u^*$\ying{. Above} $u^*$
agents rapidly commit to non-neutral opinions, a
property we exploit for vehicle coordination via \ying{zone-based} attention.

\subsection{Zone-Based Attention and Commitment Mechanism}

The intersection region is partitioned into three
zones according to each vehicle's distance from the
intersection center (which is the origin), as shown in \ying{Fig.~}\ref{fig:intersection}\ying{, where} $d_i(t) = \|\mathbf{p}_i(t)\|$ and
\begin{equation*}
\mathrm{zone}_i(t) = \begin{cases}
\text{EVOLUTION}    & d_i < R_{\mathrm{evo}},
                      \; R_{\mathrm{evo}}{=}15\,\text{m},\\
\text{DECISION}     & d_i \leq R_{\mathrm{int}},
                      \; R_{\mathrm{int}}{=} 5\,\text{m}, \\
\text{EXITED}       & d_i > R_{\mathrm{int}}.
                      
\end{cases}
\label{eq:zones}
\end{equation*}

Opinion dynamics are zone-aware through a
state-dependent attention gain
\begin{equation}
u_i(t) = u_0\!\left(\mathrm{zone}_i\right)
        + K_u\!\left(z_i - \tfrac{1}{2}\right)^{\!2},
\label{eq:attention}
\end{equation}
with $u_0{=}0.5$ (EVOLUTION), $u_0{=}0.8$
(DECISION), and $K_u{=}2.0$. As $z_i$ departs from
neutral, $u_i$ increases and accelerates opinion
formation analogous to the supercritical
bifurcation in \eqref{eq:base_nod}\ying{.}
Each vehicle maintains a discrete commitment state
\begin{equation}
\sigma_i \in \bigl\{N,\; G,\; Y,\; E\bigr\},
\label{eq:sigma}
\end{equation}
where the letters denote \textsc{negotiate}, \textsc{go},
\textsc{yield}, and \textsc{exit}.
Initially $\sigma_i {=} N$. Opinion evolves
continuously while $\sigma_i {=} N$ across both
the evolution and decision zones, with attention
gain $u_i$ increasing as the vehicle approaches
the intersection. Upon entering the decision
zone, the commitment gate evaluates FCFS priority
and feasibility, freezing opinion at commitment: it freezes at $z_i{\to}1$
on $G$  and remains frozen through \textsc{exit}, while for $z_i{\to}0$ on $Y$, the vehicle only \ying{resumes} negotiation once the conflicting vehicle exits. This commitment mechanism facilitates cooperative collision free passing, which is discussed in more detail \ying{in} Section~\ref{gate}.
\subsection{Conflict Topology Based Communication Network Graph}

For each vehicle pair $(i,j)$ we compute a crossing
conflict indicator
\begin{equation*}
K(i,j) = \begin{cases}
1 & \text{paths geometrically cross },\\
0 & \text{otherwise},
\end{cases}
\end{equation*}
and a merge indicator
\begin{equation*}
M(i,j) = \begin{cases}
1 & \text{out-lane}_i = \text{out-lane}_j,\\
0 & \text{otherwise,}
\end{cases}
\end{equation*}
 showing whether they share a common out-lane.
Both crossing and merge conflicts require sequencing;
$M$ is maintained separately to distinguish the type
of deferral in the trajectory \ying{optimization} layer. We define the static
signed adjacency matrix $A \in
\{-1,0,+1\}^{N_a \times N_a}$, where
\begin{equation}
A(i,j) = \begin{cases}
-1 & K(i,j)=1 \text{ or } M(i,j)=1 ,\\
+1 & K(i,j)=0 \text{ and } M(i,j)=0 ,\\
 0 & i = j.
\end{cases}
\label{eq:A}
\end{equation}
When $A(i,j) = -1$ we have an antagonistic edge (conflict
requires sequencing) and when $A(i,j) = +1$ we have a cooperative edge
(no conflict). Since geometric conflict is symmetric,
$A(i,j) = A(j,i)$  and $A$ is generally an
undirected signed graph, calculated using declared maneuver intentions.

\subsection{Commitment State Aware Belief System Network Graph}

The dynamic belief matrix $\mathbf{B}(t) \in
\{-1,0,+1\}^{N_a \times N_a}$ encodes the current
commitment state of each neighbor, received via V2V
\begin{equation}
B(i,j) = \begin{cases}
+1 & \sigma_j = \textsc{go},\\
-1 & \sigma_j = \textsc{yield}, \\
 0 & \text{otherwise.}
\end{cases}
\label{eq:B}
\end{equation}
Unlike $A$, the belief matrix is  directed $B(i,j) \neq B(j,i)$,
since $\sigma_i$ and $\sigma_j$ evolve independently, 
and is sparse throughout most of the scenario\ying{.} It is nonzero only when a neighbor has committed
and is actively in the decision or intersection zone.
An uncommitted neighbor ($B(i,j)=0$) exerts no
influence on $i$'s opinion, preventing undecided
vehicles from disrupting each other's opinion
trajectories.

\subsection{Proposed Decision Model for Intersection}

Each vehicle $i$ maintains a scalar opinion
$z_i \in [0,1]$, where $z_i \to 1$ represents GO,  $z_i \to 0$ YIELD , and $z_i = 0.5$
neutral. Each committed neighbor $j$ (i.e.\
$B(i,j) \neq 0$) is assigned to one of the three
interaction channels based on the joint state of
$A(i,j)$ and $B(i,j)$
\begin{align}
\mathcal{N}^-_+(i) =
  \{j : A(i,j){=}{-1},\; B(i,j){=}{+1}\} \nonumber,\\
\mathcal{N}^-_-(i) =
  \{j : A(i,j){=}{-1},\; B(i,j){=}{-1}\} \nonumber,\\
\mathcal{N}^+(i)   =
  \{j : A(i,j){=}{+1},\; B(i,j){\neq}{0}\} \nonumber.
  \label{eq:N1}
\end{align}
This is our key departure from \eqref{eq:internal}, where $a_{ik}$ is a continuous multiplier. 
Here $A$ and $B$ determine which channel a
neighbor belongs to, while independent channel gains
govern influence strength. The three channel signals
are:
\\
\textbf{Suppression} 
\begin{equation*}
\mathcal{S}_i = \max_{j\,\in\,\mathcal{N}^-_+(i)} z_j,
\label{eq:S}
\end{equation*}
where conflicting
neighbor-committed GO drives $z_i \to 0$.
\textbf{Permission}
\begin{equation*}
\mathcal{P}_i =
\operatorname{mean}_{j\,\in\,\mathcal{N}^-_-(i)}(1-z_j),
\label{eq:P}
\end{equation*}
where conflicting
neighbor-commited YIELD drives $z_i \to 1$.
\textbf{Coordination}
\begin{equation*} 
\mathcal{C}_i =
\operatorname{mean}_{j\,\in\,\mathcal{N}^+(i)} z_j,
\label{eq:C}
\end{equation*}
where cooperative
neighbor- committed GO/YIELD drives alignment.
The internal opinion state aggregating all three channels is
\begin{equation}
I_i = \alpha_s\!\left(z_i - \tfrac{1}{2}\right)
    - \alpha_S\,\mathcal{S}_i
    + \alpha_P\,\mathcal{P}_i
    + \alpha_C\,\mathcal{C}_i,
\label{eq:Ii}
\end{equation}
where $\alpha_S {>} \alpha_P {>} \alpha_C$ ensures
suppression always dominates for safety. The \ying{opinion-based} decision dynamics in \eqref{eq:base_nod} are adopted
from the range $[-1,1]$ to the range $[0,1]$ so 
\begin{equation}
\tau_z\,\dot{z}_i =
  -d\,z_i +
  \tfrac{1}{2}\!\left(1 + \tanh(u_i\,I_i)\right),
\label{eq:opinion}
\end{equation}
where $d{=}1$ and $\tau_z{=}0.1$\,s. The
$(1+\tanh(x))/2$ maps $I_i$ to $(0,1)$
without boundary saturation, and the neutral point
$z_i{=}0.5$ provides symmetric recovery after
resumption.

The opinion-like dynamics in \eqref{eq:opinion}
are inspired by signed-graph nonlinear opinion
dynamics but 
the commitment decision is determined entirely by the
predictive gate\ying{,} not by any opinion threshold crossing.

\subsection{Predictive Commitment Gate and Velocity Optimization}
\label{gate}

Rather than applying MPC as a monolithic trajectory
optimizer, we decompose its two constituent ideas:
a)~predictive constraint evaluation drives the commitment
gate, determining safe crossing order; and b)~single-step cost
minimization drives velocity regulation, ensuring smooth
speed adaptation consistent with $\sigma_i$, as explained by   \eqref{eq:sigma}. This
decomposition eliminates the need for a QP solver while
preserving the safety guarantees of predictive constraint
satisfaction.

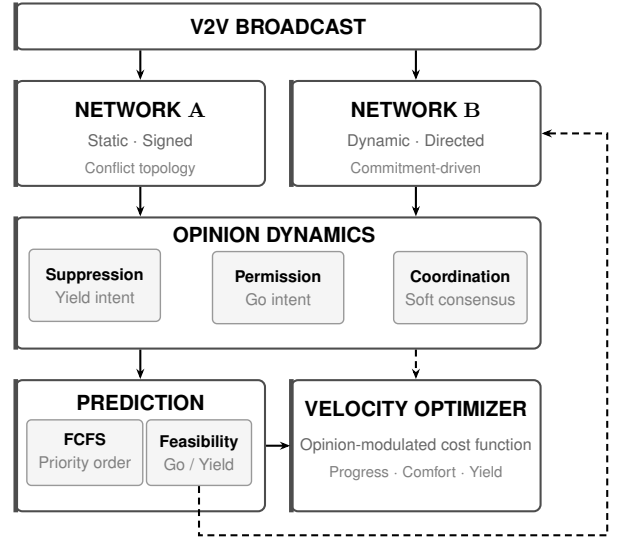
\begin{figure}[t]
\centering
\resizebox{0.96\columnwidth}{!}{%
\begin{tikzpicture}[
  node distance=0pt,
  background rectangle/.style={fill=white},
  show background rectangle,
]

\def\TW{8.0cm}
\def\HW{3.8cm}
\def\ROWGAP{0.45cm}

\node[mbox, minimum width=\TW, minimum height=0.7cm] (V2V) at (0,0) {};
\draw[black!70, line width=2pt, line cap=butt]
  ([xshift=0.5pt]V2V.north west) -- ([xshift=0.5pt]V2V.south west);
\node[ltitle=black, anchor=center] at (V2V.center) {V2V BROADCAST};

\node[mbox, minimum width=\HW, minimum height=1.6cm,
      below left=\ROWGAP and 0pt of V2V.south west,
      anchor=north west] (boxA) {};
\draw[black!70, line width=2pt, line cap=butt]
  ([xshift=0.5pt]boxA.north west) -- ([xshift=0.5pt]boxA.south west);
\node[ltitle=black, anchor=north, yshift=-6pt] at (boxA.north)
  {NETWORK $\mathbf{A}$};
\node[slabel=black, anchor=north, yshift=-20pt] at (boxA.north)
  {Static $\cdot$ Signed};
\node[tlabel=black, anchor=north, yshift=-32pt] at (boxA.north)
  {Conflict topology};

\node[mbox, minimum width=\HW, minimum height=1.6cm,
      below right=\ROWGAP and 0pt of V2V.south east,
      anchor=north east] (boxB) {};
\draw[black!70, line width=2pt, line cap=butt]
  ([xshift=0.5pt]boxB.north west) -- ([xshift=0.5pt]boxB.south west);
\node[ltitle=black, anchor=north, yshift=-6pt] at (boxB.north)
  {NETWORK $\mathbf{B}$};
\node[slabel=black, anchor=north, yshift=-20pt] at (boxB.north)
  {Dynamic $\cdot$ Directed};
\node[tlabel=black, anchor=north, yshift=-32pt] at (boxB.north)
  {Commitment-driven};

\node[mbox, minimum width=\TW, minimum height=2.0cm,
      below=\ROWGAP of boxA.south west,
      anchor=north west] (boxOp) {};
\draw[black!70, line width=2pt, line cap=butt]
  ([xshift=0.5pt]boxOp.north west) -- ([xshift=0.5pt]boxOp.south west);
\node[ltitle=black, anchor=north west, xshift=65pt, yshift=-1pt]
  at (boxOp.north west) {OPINION DYNAMICS};

\node[sbox, minimum width=2.0cm, minimum height=1.1cm,
      anchor=west, xshift=6pt, yshift=-1pt]
  (Sch) at (boxOp.west |- boxOp.center)
{
  \textbf{Suppression}\\[2pt]
  \textcolor{black!50}{Yield intent}
};

\node[sbox, minimum width=2.0cm, minimum height=1.1cm,
      anchor=center, yshift=-2pt]
  (Pch) at (boxOp.center)
{
  \textbf{Permission}\\[2pt]
  \textcolor{black!50}{Go intent}
};

\node[sbox, minimum width=2.0cm, minimum height=1.1cm,
      anchor=east, xshift=-6pt, yshift=-2pt]
  (Cch) at (boxOp.east |- boxOp.center)
{
  \textbf{Coordination}\\[2pt]
  \textcolor{black!50}{Soft consensus}
};

\node[mbox, minimum width=\HW, minimum height=2.0cm,
      below left=\ROWGAP and 0pt of boxOp.south west,
      anchor=north west] (boxGate) {};
\draw[black!70, line width=2pt, line cap=butt]
  ([xshift=0.5pt]boxGate.north west) -- ([xshift=0.5pt]boxGate.south west);
\node[ltitle=black, anchor=north, yshift=-3pt] at (boxGate.north)
  {PREDICTION};

\node[sbox, minimum width=1.5cm, minimum height=1.0cm,
      anchor=west, xshift=5pt,yshift = -3pt]
  (FCFS) at (boxGate.west |- boxGate.center)
{
  \textbf{FCFS}\\[2pt]
  \textcolor{black!50}{Priority order}
};

\node[sbox, minimum width=1.6cm, minimum height=1.0cm,
      anchor=east, xshift=-6pt, yshift=-3pt]
  (feas) at (boxGate.east |- boxGate.center)
{
  \textbf{Feasibility}\\[2pt]
  \textcolor{black!50}{Go / Yield}
};

\node[mbox, minimum width=\HW, minimum height=2.0cm,
      below right=\ROWGAP and 0pt of boxOp.south east,
      anchor=north east] (boxOpt) {};
\draw[black!70, line width=2pt, line cap=butt]
  ([xshift=0.5pt]boxOpt.north west) -- ([xshift=0.5pt]boxOpt.south west);
\node[ltitle=black, anchor=north, yshift=-6pt] at (boxOpt.north)
  {VELOCITY OPTIMIZER};
\node[slabel=black, anchor=north, yshift=-22pt] at (boxOpt.north)
  {Opinion-modulated cost function};
\node[tlabel=black, anchor=north, yshift=-34pt] at (boxOpt.north)
  {Progress $\cdot$ Comfort $\cdot$ Yield};

\draw[sig=black] (V2V.south -| boxA.north) -- (boxA.north);
\draw[sig=black] (V2V.south -| boxB.north) -- (boxB.north);
\draw[sig=black] (boxA.south) -- (boxA.south |- boxOp.north);
\draw[sig=black] (boxB.south) -- (boxB.south |- boxOp.north);
\draw[sig=black] (boxOp.south -| boxGate.north) -- (boxGate.north);
\draw[sigdash] (boxOp.south -| boxOpt.north) -- (boxOpt.north);
\draw[sig=black] (boxGate.east) -- (boxOpt.west);
\draw[sigdash]
  (feas.south) -- ++(0,-0.75)
  -| ([xshift=1cm]boxB.east) -- (boxB.east);

\end{tikzpicture}
}
\caption{Architecture of the proposed framework }
\label{fig:arch}
\end{figure}

On entry to the decision zone, vehicle $i$ records
its arrival timestamp $t_i^{\mathrm{dz}}$ and
broadcasts it via V2V communication as shown in Fig.~\ref{fig:arch} alongside the states 
$(\sigma_i, z_i, \mathbf{x}_i)$.

\noindent\textbf{Predictive feasibility check:}
Vehicle $i$ projects its earliest crossing window,
\begin{equation*}
T_i = \Bigl[t + \tfrac{d_i}{v^{\max}_i},\;\;
            t + \tfrac{d_i + L_{\mathrm{box}}}
            {v^{\max}_i}\Bigr]
\label{eq:Ti}
\end{equation*}
where $L_{\mathrm{box}} {=} 2R_{\mathrm{int}} {+}
L_v$. Feasibility requires non-overlapping occupancy
windows against all $\sigma_j {=} G$ neighbors with
$K(i,j){=}1$
\begin{equation}
\min\!\bigl(T_i(2),T_j(2)\bigr) -
\max\!\bigl(T_i(1),T_j(1)\bigr) < \varepsilon,
\quad \varepsilon{=}0.6\,\text{s}.
\label{eq:feasibility}
\end{equation}
Feasible implies $\sigma_i \to G$,
$T_i$ stored and broadcast; infeasible
implies $\sigma_i \to Y$.
For merge conflicts $M(i,j){=}1$, a
time-to-merge-point check with margin
$t_{\mathrm{margin}}{=}1.5$\,s replaces
\eqref{eq:feasibility}.

Each commitment event updates $B(j,i)$ for all
neighbors via V2V, activating the channels in \ying{~}\eqref{eq:opinion} and
inducing an opinion spike among required vehicles, either  $z_j{\to}0$ or
$z_j{\to}1$ mimicking\ying{~}\cite{10817538}. 
\\

\noindent\textbf{Single-Step Multi-Objective Velocity Optimization}

Each vehicle follows its pre-computed geometric path
exactly via arc-length parameterization. The Optimization
layer determines only longitudinal acceleration $a_i^*$
at each timestep solving a multi objective cost function modulated by opinion,
\begin{mini*}|l|
  {a \in \mathcal{A}}
  {J_i(a) = J_{\mathrm{prog}} + J_{\mathrm{comf}}
           + J_{\mathrm{spat}} + J_{\mathrm{yield}}}
  {\label{eq:opt}}{}
  \addConstraint{a}{\in [-5.0,\; 2.5]\;\text{m/s}^2}{}
\end{mini*}
where $\mathcal{A}$ is a uniform grid of $N_c{=}15$
candidates over the acceleration limits. And
$v^{\max}_i \in \{11.1,\,8.0,\,7.0\}$\,m/s is the speed limit for
straight, left and right maneuvers respectively, ensuring safe cornering speeds across all
maneuver types. 

\noindent\textbf{a) Progress Cost:}
\begin{equation*}
J_{\mathrm{prog}} = w_p^{(\sigma,z)}\,
\frac{d_i}{\max(v_{\mathrm{pred}},\,0.1)}.
\end{equation*}
The predicted velocity is
clipped to a maneuver-dependent speed limit
\begin{equation*}
v_{\mathrm{pred}} = \mathrm{max}
    (0,min(v_i {+} a\Delta t, v^{\max}_i)),
\end{equation*}
$v_i^{\max}$ reflects the geometric
constraint of the assigned maneuver and
\vspace{0em}
\begin{equation*}
w_p^{(\sigma,z)} =
\begin{cases}
10\,w_p & \sigma_i{=}G,\;
           d_i \leq R_{\mathrm{int}},\\
(0.5 + z_i)\,w_p & \sigma_i{=}N,\\
w_p & \text{otherwise.}
\end{cases}
\label{eq:wprog}
\end{equation*}
In the negotiating state, $z_i \to 1$ (GO)
increases approach aggressiveness while $z_i \to 0$
(YIELD) reduces it, providing a smooth
pre-commitment speed adaptation. Where $w_p$ = 1.

\noindent\textbf{b) Comfort Cost:}
Acceleration effort is penalized quadratically
to suppress aggressive maneuvers
\begin{equation*}
J_{\mathrm{comf}} = w_c\, a^2,
\label{eq:Jcomf}
\end{equation*}
where $w_c {=} 0.5$. 

\noindent\textbf{c) Spatial Repulsion Cost:} from conflicting
neighbors $K(i,j){=}1$
\begin{equation*}
J_{\mathrm{spat}} = \!\sum_{j\,\in\,{K}_i}
w_j\;\exp\!\bigl(-{\tfrac{1}{2}}(d_{ij}
- d_{\mathrm{safe}})\bigr),
\label{eq:Jspat}
\end{equation*}
\vspace{-1em}
\begin{equation*}
w_j =
\begin{cases}
w_{\mathrm{com}}
  & \sigma_j{=}G,\\
w_{\mathrm{dec}}\,(0.5 + z_j)
  & \sigma_j{=}N,\; d_j < R_{\mathrm{evo}},\\
w_{\mathrm{evo}}
  & \text{otherwise.}
\end{cases}
\label{eq:wj}
\end{equation*}
A negotiating neighbor leaning GO ($z_j{\to}1$)
induces stronger repulsion than one leaning YIELD
($z_j{\to}0$), reflecting its expressed crossing
intent before commitment.
$J_{\mathrm{spat}}$ is suppressed for crossing
conflicts when $\sigma_i{=}G$ and
$d_i \leq R_{\mathrm{int}}$, where $w_{com}$ = 1000, $w_{dec}$ = 10, $w_{evo}$ = 1 and the safe distance $d_{safe}$ = 3m.

\noindent\textbf{d) Yield Braking Cost:}
When $\sigma_i{=}Y$,
a braking cost is applied with exponential onset
beyond $R_{\mathrm{int}}
\,\text{}$ ensuring smooth deceleration,
\begin{equation*}
J_{\mathrm{yield}} =
\begin{cases}
w_{\mathrm{com}}\,v_{\mathrm{pred}}
  & d_i \leq R_{\mathrm{int}},\\[2pt]
0.05\,w_{\mathrm{com}}\,v_{\mathrm{pred}}\,
  e^{-(d_i - R_{\mathrm{int}})/6}
  & d_i > R_{\mathrm{int}},
\end{cases}
\label{eq:Jyield}
\end{equation*}
avoiding
discontinuous braking at the stop boundary.
 
\section{Simulation Results and Analysis}
We validate the proposed framework in MATLAB across all 81 combinatorial maneuver scenarios involving 4 CAVs approaching a 4-way unsignalized intersection, achieving improved performance over FCFS across all configurations. While the all-right-turn case gives a fully cooperative signed network with trivially parallel passing and no sequencing challenge, 3 representative cases are presented as in Table~\ref{tab:results}.

\begin{figure}[H]
\centering
\includegraphics[width=\columnwidth]{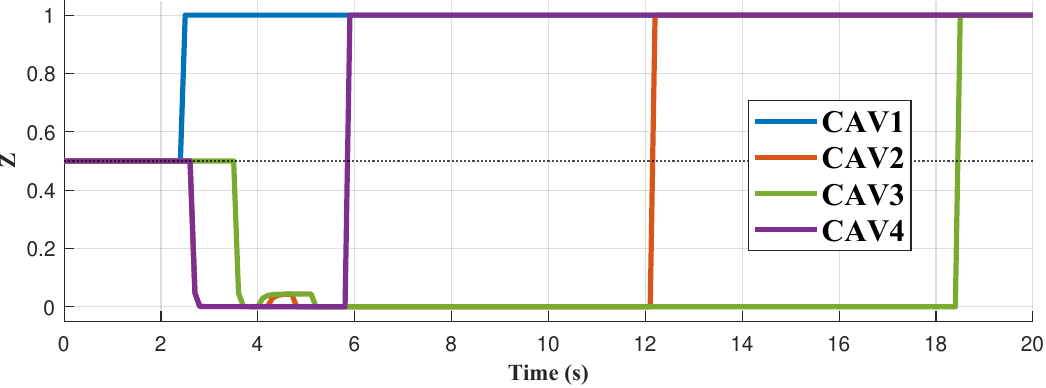}
\includegraphics[width=\columnwidth]{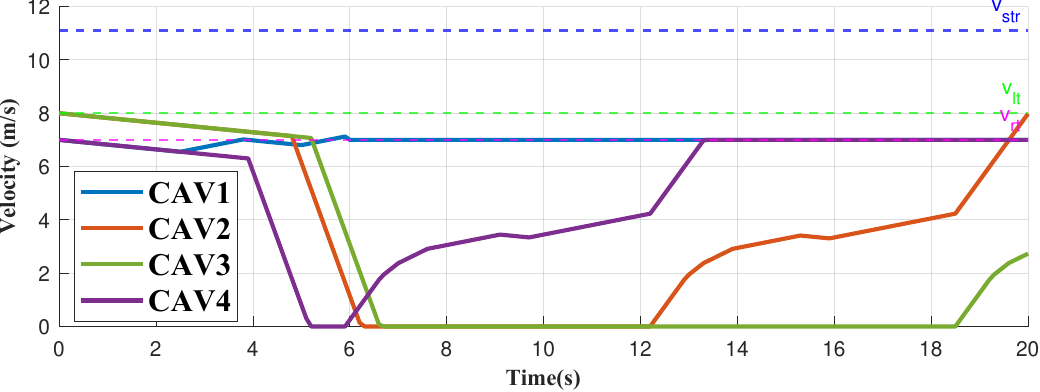}
\caption{Scenario 1 (all left turns): (a) $z_i$;
(b)  speeds}
\label{fig:config22}
\end{figure}

\begin{figure}[H]
\centering
\includegraphics[width=0.75\columnwidth]{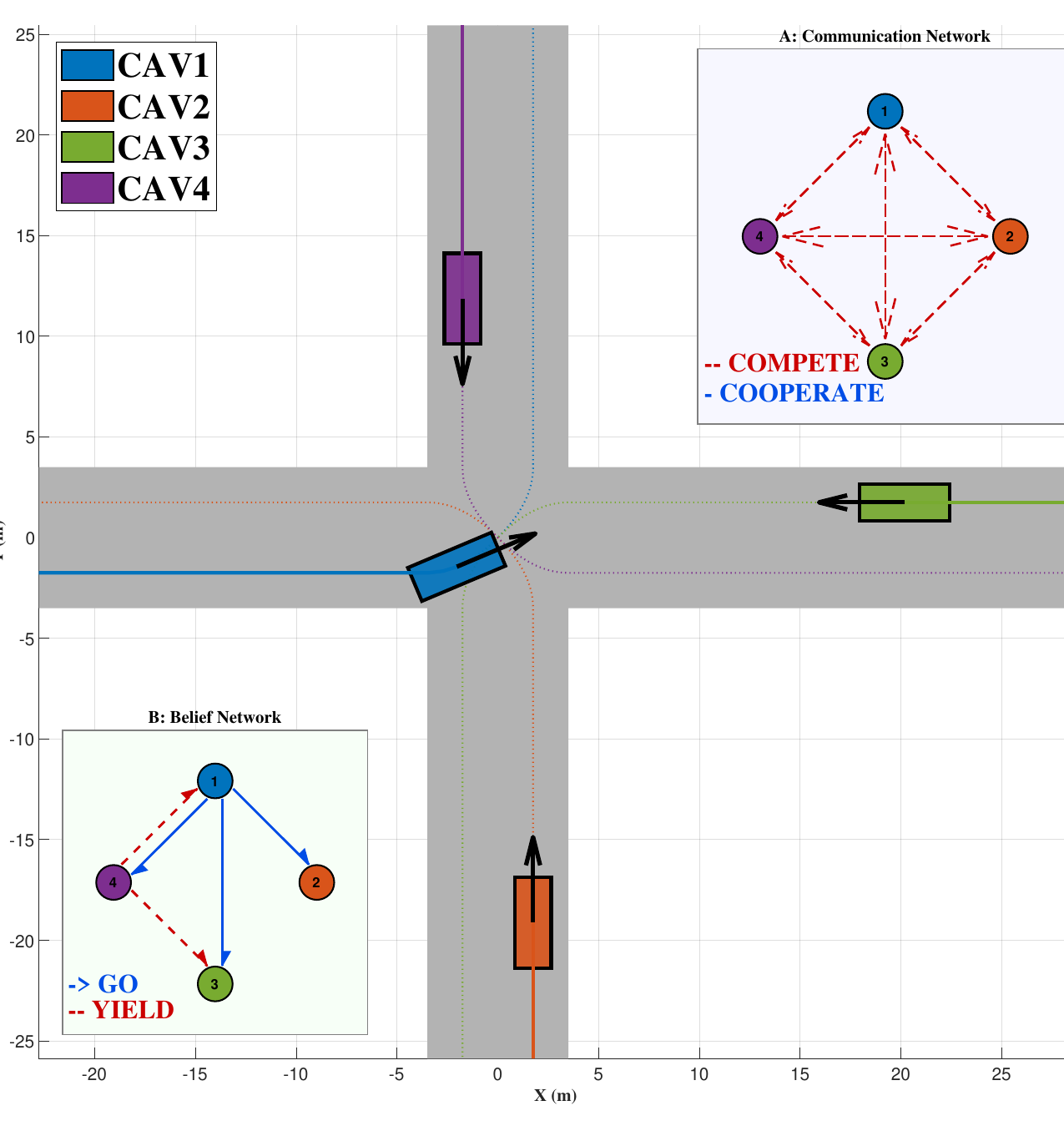}
\includegraphics[width=0.75\columnwidth]{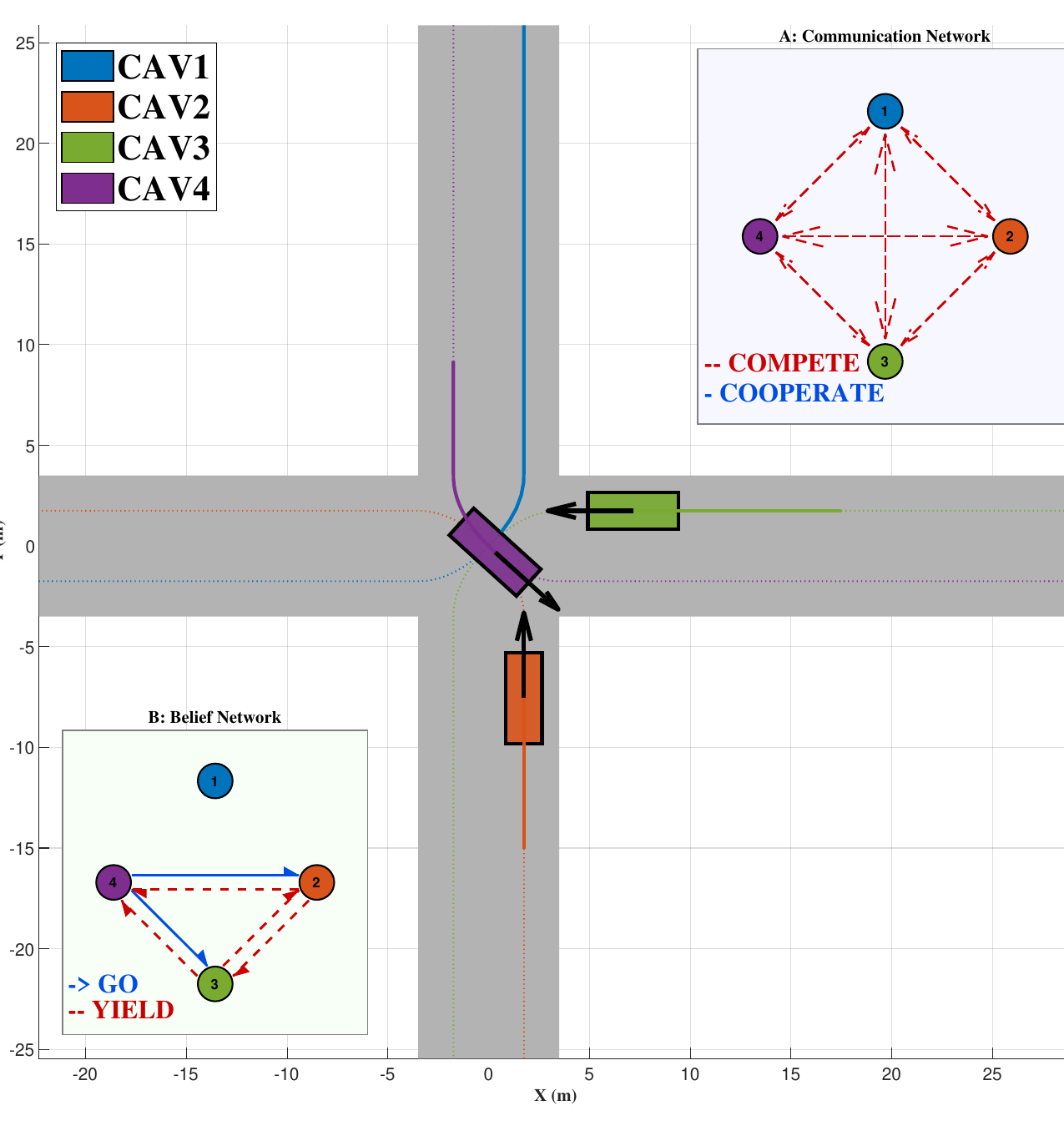}
\caption{Scenario 1(all left turns) snapshots: 4s and 6.5s}
\label{fig:config2}
\end{figure}

Scenario 1 introduces maximum contention: all 4
vehicles intend left turns, every pairwise edge in 
$A$ is competitive (-1), 
Vehicles resolve priority through FCFS 
ordering based on decision zone arrival time, with opinions evolving 
accordingly under full suppression channel activation. As shown in 
Fig.~\ref{fig:config2} and the opinion and speed profiles in 
Fig.~\ref{fig:config22}, CAV1 is granted priority first, followed 
by CAV3, then CAV2, and finally CAV4, each commitment propagating 
through $B$ and pre-conditioning the subsequent vehicle's 
opinion transition.  All vehicles traverse the intersection 
conflict-free, demonstrating that the signed network structure 
resolves fully competitive topologies without any explicit sequencing 
constraint.
 Scenario 2 introduces complex conflict topology, comprising 
two merge conflicts and one crossing conflict under mixed maneuver 
intentions. FCFS priority is respected as the default ordering, yet an emergent reordering arises when a feasible gap opens while CAV4 
yields for CAV1, CAV2 identifies the gap and commits to GO breaking strict FCFS order without any explicit 
reordering logic.
 
\begin{figure}[H]
\centering
\includegraphics[width=\columnwidth]{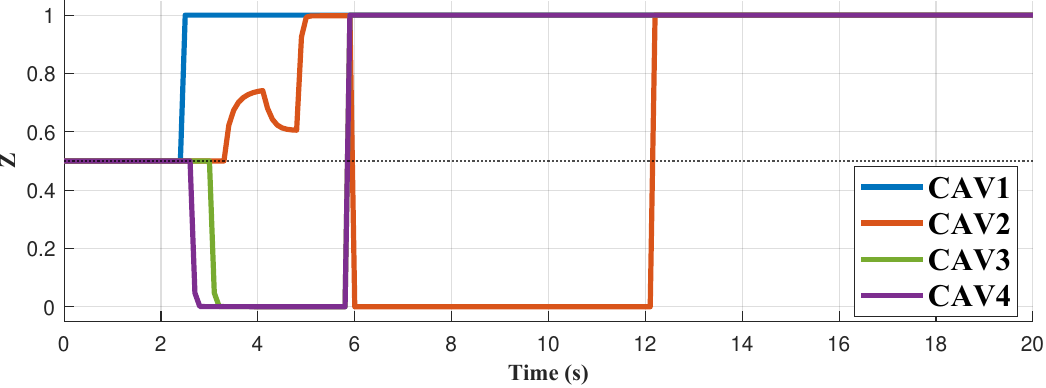}
\includegraphics[width=\columnwidth]{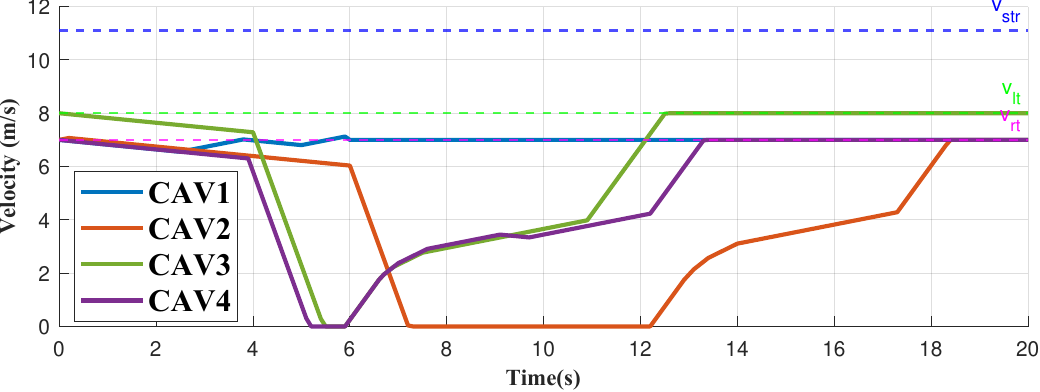}
\caption{Scenario 2(mixed manuevers): (a) $z_i$;
(b) speeds.}
\label{fig:config33}
\end{figure}

\begin{figure}[H]
\centering
\includegraphics[width=\columnwidth]{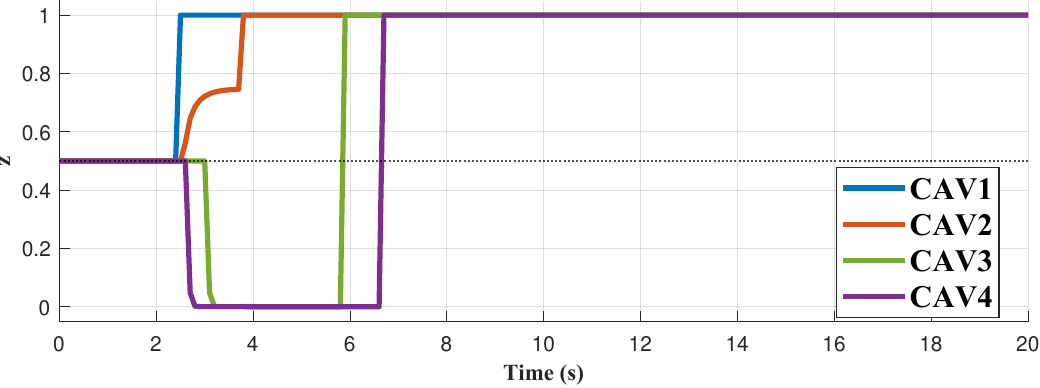}
\includegraphics[width=\columnwidth]{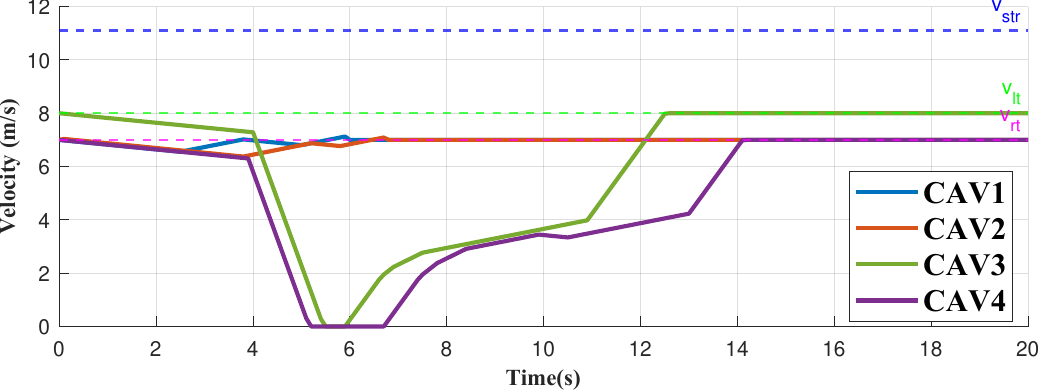}
\caption{Scenario 3(mixed maneuvers): (a) $z_i$;
(b) speeds.}
\label{fig:config44}
\end{figure}

 This optimal order emerging is captured in Scenario 3 which  is run under slightly perturbed initial position of CAV2, with the resulting opinion trajectories shown in  
Fig.~\ref{fig:config33} and 
Fig.~\ref{fig:config44} clearly showing a jump in decision making. 
\begin{figure}[H]
\centering
\includegraphics[width=0.78\columnwidth]{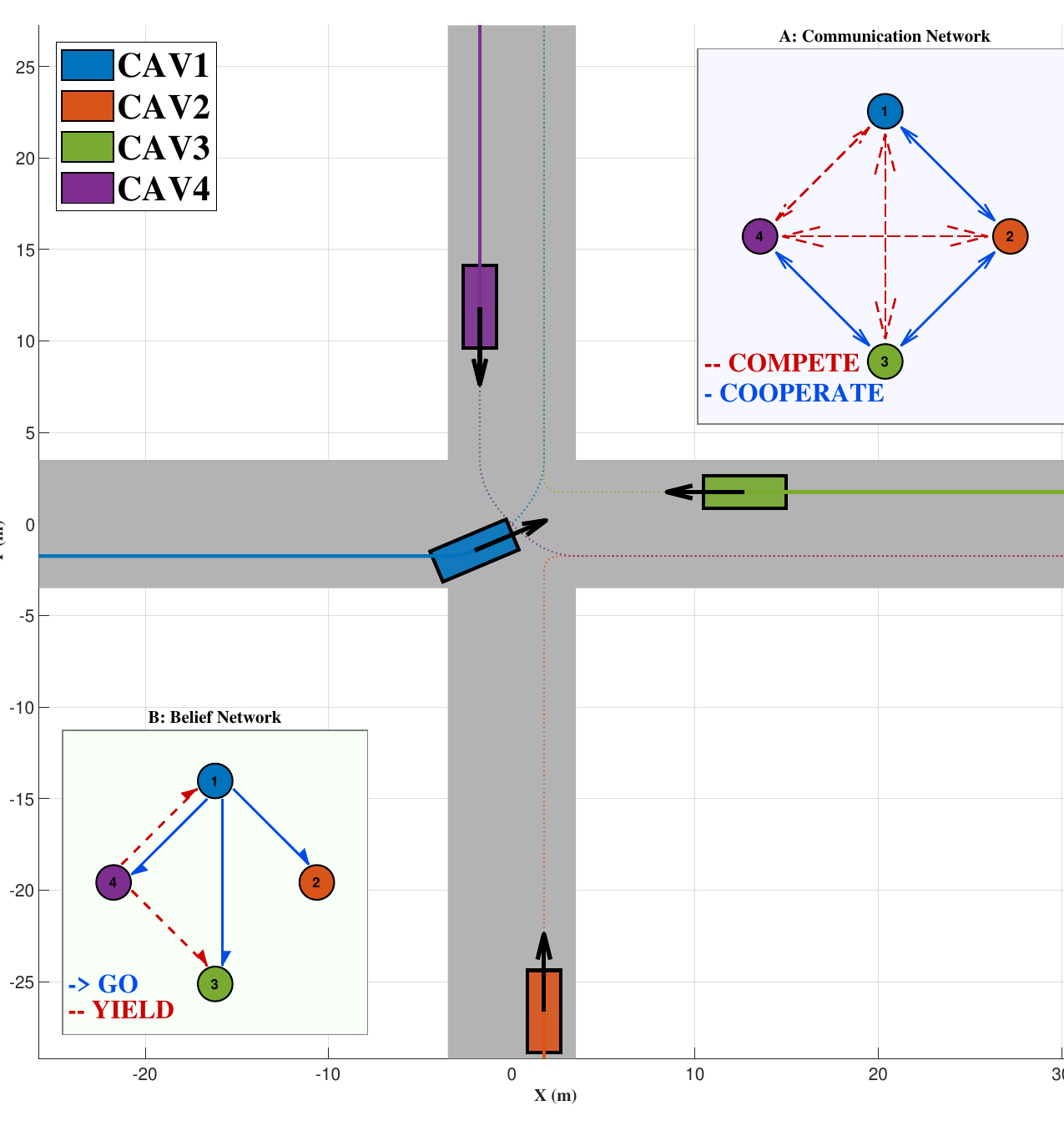}
\includegraphics[width=0.78\columnwidth]{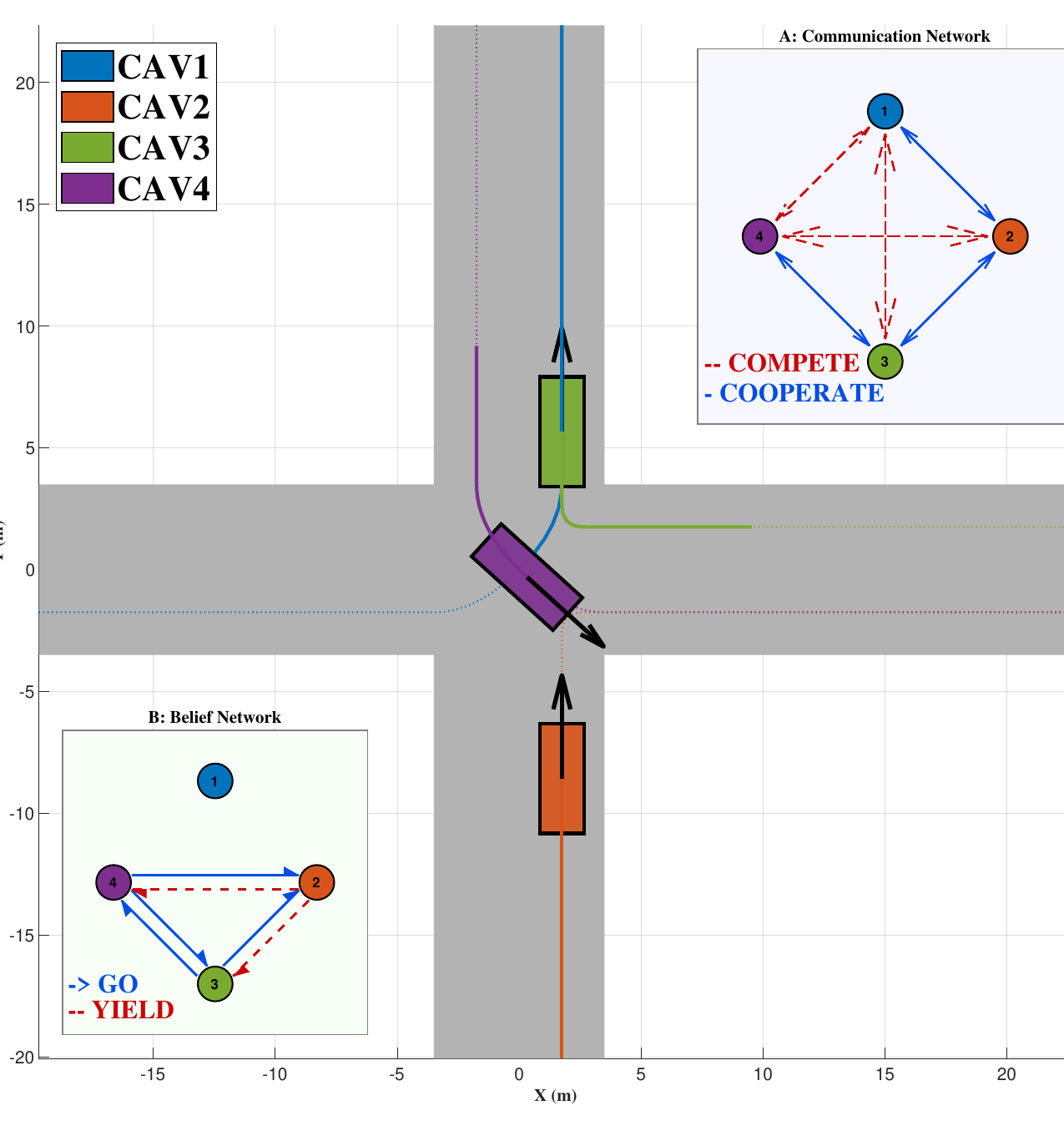}
\caption{Scenario 2 (mixed) snapshots: 4s and 9.5s }
\label{fig:config3}
\end{figure}

\begin{table*}[!t]
\centering
\setlength{\tabcolsep}{18pt}
\caption{Scenario configurations and Last vehicle exit times (s), where  CAV Entry: (in-lane, out-lane, $d_i$(m), $v_i$(m/s))}
\label{tab:results}
\begin{tabular}{ccccccc}
\toprule
Scenario & CAV1 & CAV2 & CAV3 & CAV4 & FCFS (s) & Proposed. (s) \\
\midrule
1 & (1, 6, 29, 7) & (3, 8, 45, 8) & (5, 2, 50, 8) & (7, 4, 35, 7) & 22.20 & 22.25 \\
2 & (1, 6, 29, 7) & (3, 4, 45, 7) & (5, 6, 50, 8) & (7, 4, 35, 7) & 17.09 & 17.20  \\
3 & (1, 6, 29, 7) & (3, 4, 44, 7) & (5, 6, 50, 8) & (7, 4, 35, 7) & 17.09 & 12.50  \\
\bottomrule
\end{tabular}

\end{table*}

\begin{figure}[H]
\centering
\includegraphics[width=0.80\columnwidth]{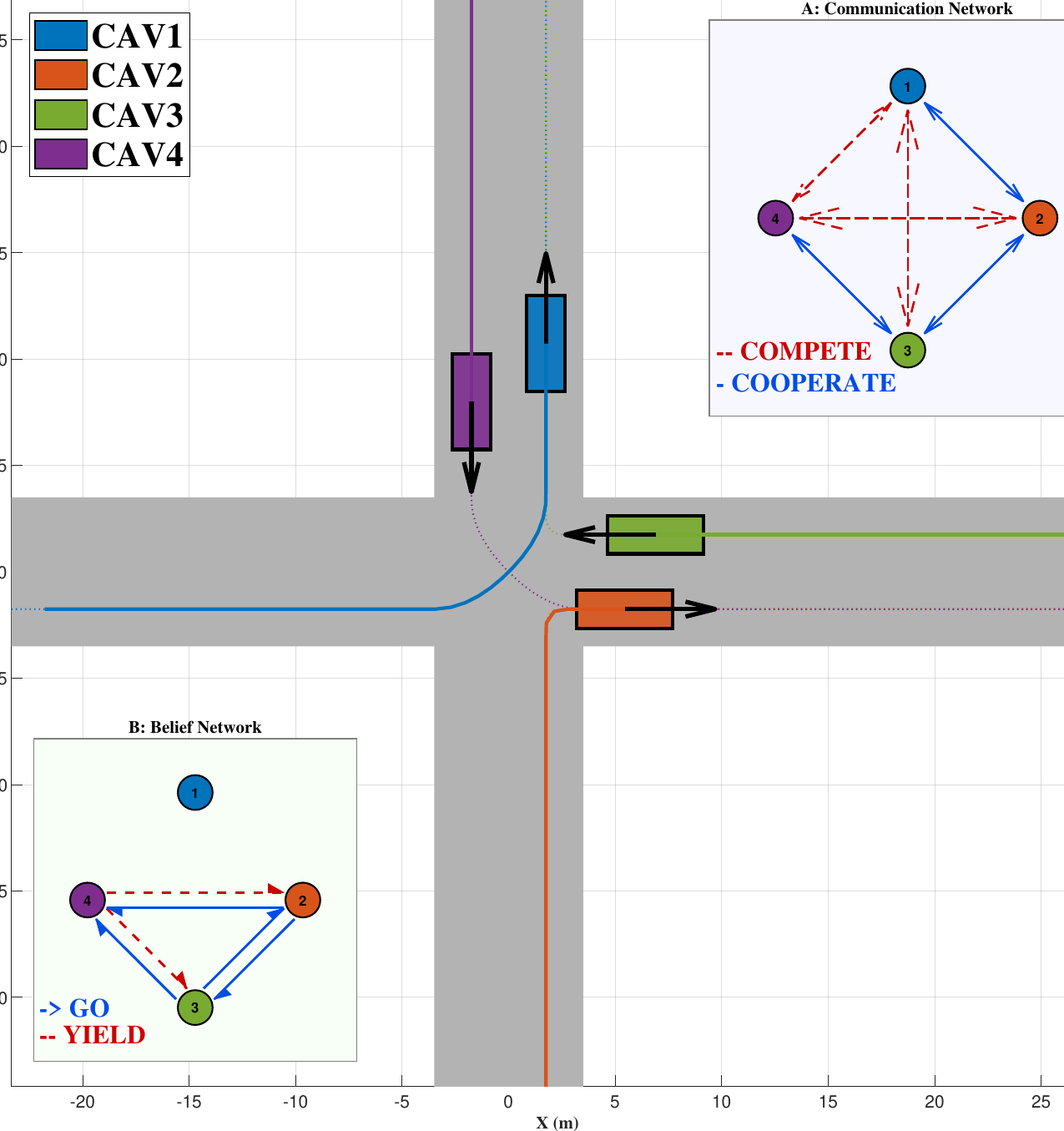}
\caption{Scenario 3(mixed) snapshot at 6s }
\label{fig:config4}
\end{figure}

\normalsize
\subsection{Analysis}
Table~\ref{tab:results} summarizes the last vehicle exit time across scenarios. In Scenario 1 and Scenario 2, the proposed method approximately equals the FCFS, consistent with the geometric structure of that scenario, where FCFS ordering is near-optimal and the opinion dynamics converge to the same sequence. While the table shows almost equal performance, it is worth noting that its achieved in a decentralized manner. Also, Scenario 3  shows how our method is adaptably efficient using scenario specific requirements and outperforms FCFS in every such interactive scenario.

\section{Conclusion}
\label{sec:conclusion}

In this letter, we address the problem of decentralized intersection coordination through a signed-graph opinion dynamics framework, where a conflict topology network and a dynamic belief network jointly drive each vehicle to a \textsc{GO} or \textsc{YIELD} commitment without a central coordinator or solver. The formulation outperforms centralized approaches by allowing crossing order to emerge from geometric feasibility and arrival priority across crossing, merge, and mixed conflict scenarios through a single unified parameterization based solely on vehicle maneuver intentions.  Future work will focus on extending the framework to incorporate uncertainty-aware belief updates, and validating the approach in mixed-traffic environments with higher density.


\bibliographystyle{IEEEtran}
\bibliography{cdc}

\end{document}